\documentclass[a4paper,11pt]{article}
\usepackage{jcappub} 
\usepackage{lineno}

\title{\boldmath Spiral structure from the interference of gravitational eigenmodes in ultralight dark matter halos}

\author[1]{Jacob Reinach\note{Corresponding author.}}
\author[2]{Hubert Bray}
\affiliation[1]{Department of Physics, Duke University, Durham, NC 27708, USA}
\affiliation[2]{Department of Mathematics, Duke University, Durham, NC 27708, USA}

\emailAdd{jacob.reinach@duke.edu}

\abstract{We use a variety of analytic and numerical techniques to approximate the geometry of individual dark matter halos when modeled as collections of ultralight scalar bosons (ULDM), followed by an analysis on the future directions and implications of this work. Ultimately, we aim to understand the morphology and creation of complex structures (like spirals) in galaxies from a geometric standpoint and to investigate the connection between dark matter and baryonic matter in galaxies. We find that superpositions of gravitational eigenmodes for ULDM in a fixed background potential can yield spiral density perturbations which rotate on the order of one hundred million years and appear within ten to twenty kpc of the halo center, both features comparable to realistic galaxies, and that this model provides a natural explanation for the density waves necessary to seed baryonic spirals.}

\begin{document}
\maketitle
\flushbottom

\section{Introduction}
One of the most prominent questions of modern physics regards the nature of dark matter. A vague description of the source of unknown gravity permeating the Universe, dark matter is responsible for a variety of astrophysical phenomena and comprises about 27\% of the mass-energy of the universe (normal
baryonic matter and radiation make up about 5\%, while dark energy makes up 68\%)\cite{ostriker2003,farnes2018}. Dark matter was first theorized as an explanation for the velocity dispersion of galaxies in clusters in the 1930's, and evidence for dark matter was reinforced by observations of the high velocities of stars in rotating disk galaxies (flattened rotation curves) in the 1970's. Dark matter has since helped explain gravitational lensing events, large-scale structure, the cosmic microwave background (CMB), and observations of phenomena like the Bullet Cluster\cite{zwicky2009,rubin1978,bergstrom2012,bertone2005,robertson2016}. 

While dark matter candidates like weakly-interacting massive particles (WIMPs) and sterile neutrinos remain theoretically viable\cite{roszkowski2018,bertone2010}, recent research has placed an emphasis on wave-like candidates of dark matter, often pseudo-scalar Nambu-Goldstone bosons like the QCD axion and ultralight dark matter (ULDM)\cite{peccei1977,ferreira2021,hui2021,hui2017}. These particles have distinct phenomenologies compared to other dark matter candidates and have a unique ability to form solitonic halo cores and density waves\cite{burkert2020,bray2010}.

One of the best ways to observationally search for dark matter utilizes the fact that dark matter forms halos. Observational techniques have proven that galaxies and galaxy clusters are surrounded by large, clumpy over-densities of dark matter called dark matter halos; however, it remains nearly impossible to quantify the exact structure of individual halos given the current status of our technology\cite{allgood2006,frenk1988}. Thus, the influence of individual dark matter halos on galaxy morphology remains an active area of theoretical research. 

One of the open questions in galaxy morphology which may be addressed through a better understanding of individual dark matter halo dynamics is the formation of long-lived spiral patterns in galaxies. Though a variety of processes like rotational shearing, gravitational instabilities after mergers, and the rotation of bar-like galaxy cores can impart spiral structure on galaxies\cite{donghia2013,goldreich1965,lieb2022}, the mechanism of formation of spirals in old, isolated galaxies remains speculative\cite{sellwood2022}. Density-wave theory suggests that asymmetric density waves permeating galaxies are able to impart dynamic, long-lived spirals in the baryonic matter we observe. However, the mechanism through which these density waves arise has not been established concretely. Wave-like dark matter, particularly ULDM in the case of this research, naturally seeds these density waves, and can provide potential insights into the relation between galaxy morphology and individual dark matter halos\cite{bray2010}. 

In this work, we model ULDM as classical field constructed out of gravitational eigenmodes in a fixed background potential. The interference of such eigenmodes in a variety of linear superpositions yield long-lived, asymmetric patterns in the reconstructed, rotating halos. While nearly all superpositions that include high-energy states display clear density waves, a number of these results also reveal spiral patterns which alternate in leading and trailing arms over time, suggesting that dark matter halos made of ULDM may help seed long-lived spiral patterns in visible galaxies.

\section{Methods}

ULDM fields satisfy the Einstein-Klein-Gordon system (for further reading on the Einstein-Klein-Gordon system, see \cite{wald1984}). Following \cite{bray2010,bray2015}, we simplify the Einstein-Klein-Gordon system using a fixed background potential that fits the simplified structure of realistic galaxy rotation curves, creating a modified Klein-Gordon equation. Because ULDM halos are single coherent fields, we model them as superpositions of the gravitational eigenmodes that satisfy our modified Klein-Gordon equation.

\begin{figure}[htbp]
    \centering
    \includegraphics[width=1\linewidth]{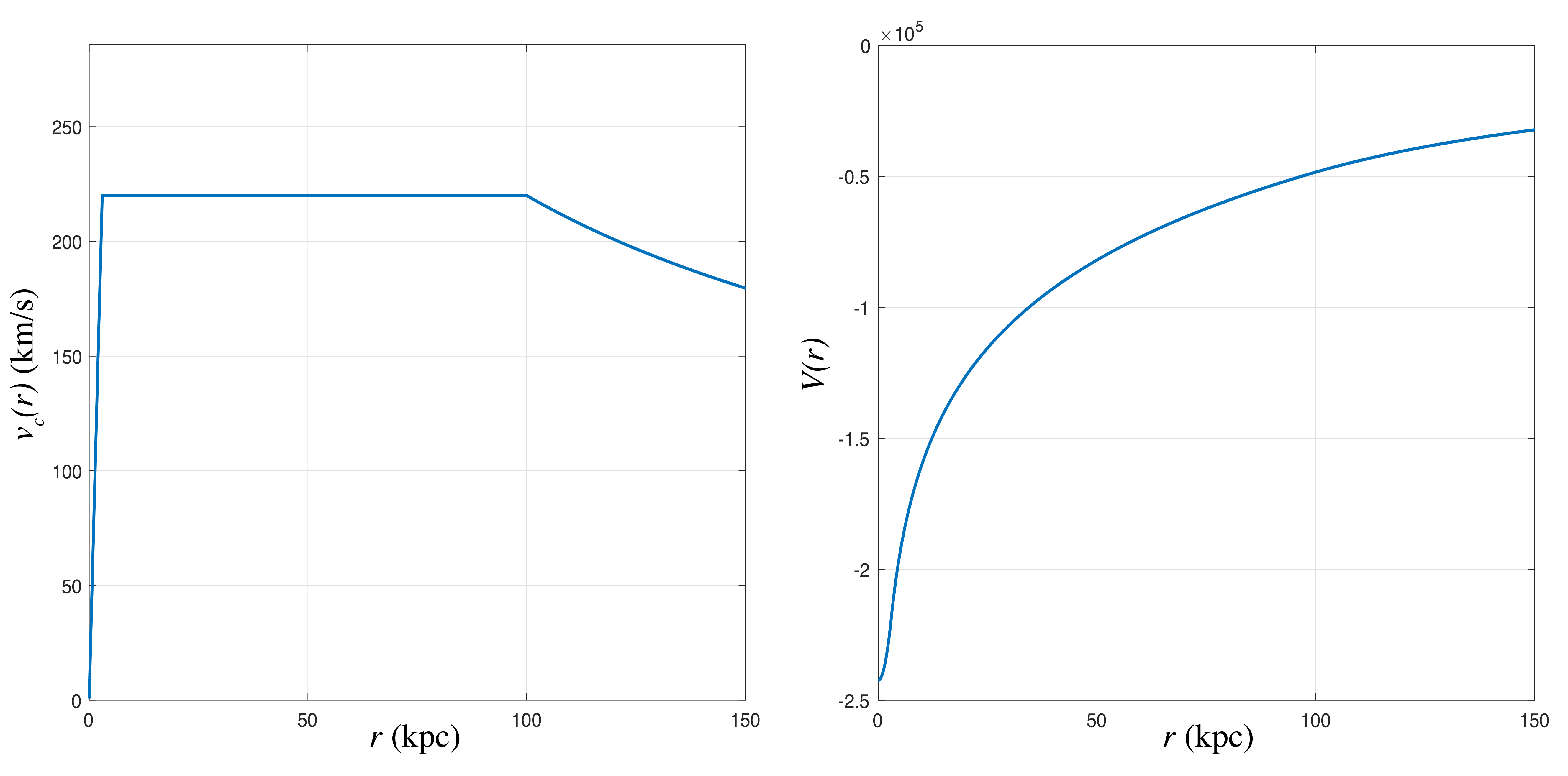}
        \caption{The image on the left is the simplified piecewise galaxy rotation curve based on observational data which shows star velocity as a function of radius from galactic center. The image to the right is the corresponding piecewise galaxy potential for this specific rotation curve. Both are phenomenological models of Vera Rubin's galaxy rotation curves rather than precision fits. We set $r_c = 3.5$ kpc, $r_f = 100$ kpc, $v_0 = 220$ km/s.}
    \label{fig:rotcurve}
\end{figure}

First, we rewrite the Klein-Gordon equation in a static, spherically symmetric spacetime signature, denoted $g_{\mu\nu}$ in eq. 2.1, 
\begin{equation}
g_{\mu\nu} =
\begin{pmatrix}
- e^{2V(r)} & 0 & 0 & 0 \\
0 & \dfrac{1}{\phi(r)} & 0 & 0 \\
0 & 0 & r^{2} & 0 \\
0 & 0 & 0 & r^{2}\sin^{2}\theta
\end{pmatrix}.
\end{equation}
We recover a Schwarzschild-like metric, where the term $g_{rr} = \frac{1}{\phi(r)}$ simplifies to $(1-\frac{2M(r)}{r})^{-1}$ in the nonrelativistic limit but is algebraically simpler to manipulate. We can then rewrite the Klein-Gordon equation (eq. 2.2) for this given metric, yielding eq. 2.3:
\begin{equation}
(\Box - m^2)\Phi = 0,
\end{equation}

\begin{equation}
\begin{aligned}
&- e^{-2V(r)}\,\partial_t^{2}\Phi
+ \phi(r)\,\partial_r^{2}\Phi
+ \left[
\phi(r)V'(r)
+ \frac{2\phi(r)}{r}
+ \frac{1}{2}\phi'(r)
\right]\partial_r \Phi \\[6pt]
&\quad + \frac{1}{r^{2}}
\left(\partial_\theta^{2}\Phi
+ \cot\theta\,\partial_\theta \Phi\right)
+ \frac{1}{r^{2}\sin^{2}\theta}\,\partial_\varphi^{2}\Phi
- m^{2}\Phi = 0 .
\end{aligned}
\end{equation}
Using the linearity of our scalar field variable $\Phi$ and the fact that we are working in a spherically symmetric, static spacetime, we can use separation of variables to simplify the field variable as shown in eq. 2.4,
\begin{equation}
\Phi(t,r,\theta,\varphi)
= e^{-i\omega t}\, R(r)\, Y_{\ell}^{m}(\theta,\varphi).
\end{equation}
Using the known eigenvalue relations which correspond to both the complex exponential and spherical harmonic terms, we can simplify eq. 2.3 into the eigenvalue equation in eq. 2.5, 
\begin{equation}
\begin{aligned}
\phi(r)\,R''(r)
&+ \left[
\phi(r)V'(r)
+ \frac{2\phi(r)}{r}
+ \frac{1}{2}\phi'(r)
\right] R'(r) \\[6pt]
&+ \left[
\omega^{2} e^{-2V(r)}
- \frac{\ell(\ell+1)}{r^{2}}
- m^{2}
\right] R(r) = 0.
\end{aligned}
\end{equation}
In order to solve this equation as a second-order ordinary differential equation, we must know the functions $V(r)$ and $\phi(r)$ explicitly, or the gravitational potential field and its corresponding spatial coordinate transformation respectively. Using the shape of galaxy rotation curves constructed in \cite{rubin1978}, we note a simple mathematical trend. Namely, from the galactic center out to a small, core radius there is a steep linear relation between stellar velocity and radius, followed by a flattened star speed until some Keplerian fall-off at sufficiently large radii. The velocity of stars can then be modeled as the piecewise function, 
\begin{equation}
v(r) =
\begin{cases}
\displaystyle \alpha r, & 0 \le r < r_c, \\[6pt]
\displaystyle v_0, & r_c \le r < r_f, \\[6pt]
\displaystyle v_0 \sqrt{\frac{r_f}{r}}, & r \ge r_f,
\end{cases}
\end{equation}
for any choice of core radius, $r_c$, outer radius of the galaxy, $r_f$, and maximum stellar velocity, $v_0$, where $\alpha = v_0 / r_c$. These velocities are plotted on the left in figure~\ref{fig:rotcurve}. Now, we simply integrate the quantity $\frac{v(r)^2}{r}$ with respect to radius in order to find the gravitational potential, $V(r)$, since $a = \frac{dV}{dr} = \frac{v(r)^2}{r}$ for circular velocities. Imposing continuity at both $r_c$ and $r_f$, as well as the boundary conditions which force the field to decay as $r\to\infty$, we find the piecewise potential in eq. 2.7 and plotted on the right in figure~\ref{fig:rotcurve}:
\begin{equation}
V(r) =
\begin{cases}
\displaystyle
\tfrac{1}{2}\frac{v_0^2}{r_c^2}\,r^2
- v_0^2\!\left[1 + \ln\!\left(\frac{r_f}{r_c}\right) + \tfrac{1}{2}\right],
& 0 \le r < r_c, \\[10pt]
\displaystyle
v_0^2 \ln\!\left(\frac{r}{r_c}\right)
+ \tfrac{1}{2}v_0^2
- v_0^2\!\left[1 + \ln\!\left(\frac{r_f}{r_c}\right) + \tfrac{1}{2}\right],
& r_c \le r < r_f, \\[10pt]
\displaystyle
-\,\frac{v_0^2 r_f}{r},
& r \ge r_f.
\end{cases}
\end{equation}

As previously mentioned, the term $\phi(r)$ in our spacetime metric is actually equivalent to $1-\frac{2M(r)}{r}$ in the nonrelativistic limit, where $M(r)$ is the mass enclosed within a given radius in spherical symmetry. For this system, we can recall Poisson's equation, $\nabla^2 V = 4\pi \rho$, which relates the gravitational potential in a spherically symmetric system to mass-density. Given the spherical symmetry of our problem, we can define the enclosed mass as $\frac{dM}{dr} = 4\pi r^{2}\rho$ and Poisson's equation becomes $\frac{1}{r^2}\frac{d}{dr}\!\left(r^2 V'(r)\right) = 4\pi \rho(r)$. Thus, 
\begin{equation}
\frac{d}{dr}\!\left(r^2 V'(r)\right)
=
\frac{dM}{dr}.
\end{equation}
Integrating this relation we can now solve for $M(r)$ and write an explicit form of $\phi(r)$ such that it is consistent with any choice of $V(r)$. We find $r^2 V'(r) = M(r)$ such that 
\begin{equation}
\phi(r)=1-2r\,V'(r). 
\end{equation}
Finally, we substitute in our given potential to solve for
\begin{equation}
\phi(r)=
\begin{cases}
\displaystyle 1-2\alpha^2 r^2, & 0\le r<r_c \\[10pt]

\displaystyle 1-2v_0^2, & r_c\le r<r_f \\[10pt]

\displaystyle 1-\dfrac{2v_0^2 r_f}{r}, & r\ge r_f
\end{cases}.
\end{equation}
Having written explicit forms of $V(r)$ and $\phi(r)$, we can solve for the energy eigenvalue $\omega$ in eq. 2.5 for any angular momentum number $\ell$ and unitless mass, $m=1$, which is scaled later, along with any corresponding parameters. We solve for the energy eigenvalues of the gravitational eigenmodes as a shooting problem, and find the solutions of both ground and excited states (up to $n = 5$) for all $\ell=0,1,2,3,4$. Within these numerics, we enforce $R(r) \propto r^{\ell}$ as $r \to 0$ to ensure regularity, and $R(r) \to 0$ as $r \to \infty$ for the physicality of bound states. The correct solutions for the eigenfrequency $\omega$ are such that the function decays at large $r$ rather than diverges. The resulting radial functions and their corresponding $\ell, \omega$ values get plugged into eq. 2.4 to reconstruct a full basis mode. For this specific research, we chose to examine the maximally rotating spherical harmonics such that $m = \pm \ell$, though future research should test other magnetic quantum numbers. We now write each basis field as, 
\begin{equation}
\Phi_{n\ell m}(t,r,\theta,\phi)
=
R_{n\ell}(r)\,e^{-i\omega_{n\ell} t}\,
Y_\ell^{\pm \ell}(\theta,\phi).
\end{equation}
In order to build a coherent field representative of a ULDM halo, we take an arbitrary superposition of these bases, with $A_{n\ell m}$ acting as a normalized scaling factor: 
\begin{equation}
\Psi(t,r,\theta,\phi)
=
\sum_{n,\ell,m}
A_{n\ell m}\,
\Phi_{n\ell m}(t,r,\theta,\phi)\
\end{equation}
The dynamics of interest appear when we investigate the mass density of our dark matter field, $\rho_{DM} = |\Psi|^2 = \Psi^* \Psi$. The normalization of this field obeys the relation $M_{\mathrm{halo}} = m \sum_{n,\ell,m} |A_{n\ell m}|^2$. Since individual eigenmodes are mutually orthonormal, this condition sums over strictly positive terms without cross-term interference, and thus conserves the occupation number and mass of the dark matter halo. It is worth noting that our model is limited to individual dark matter halos; halo mergers, where basis states are solved in different gravitational potentials so are not mutually orthogonal, necessitate the adaptation of this framework into a new basis. 

\section{Results}

\subsection{Eigenspectrum}
\begin{figure}[htbp]
    \centering
    \includegraphics[width=0.75\linewidth]{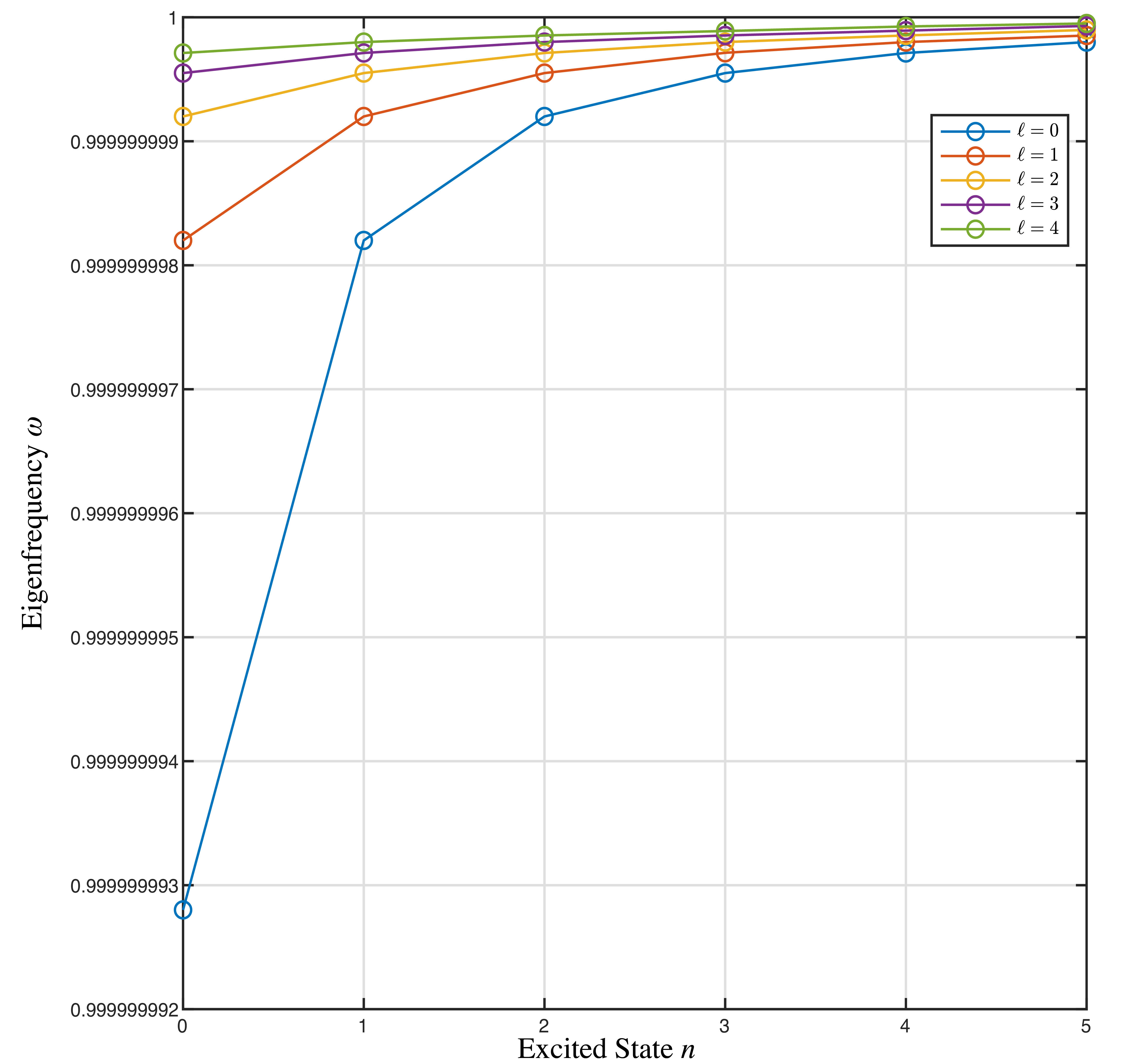}
        \caption{This figure shows the distinct eigenvalues of the radial KG equation for different angular momenta values. We use $v_0 = 0.001c$, $r_c = 5$ kpc, $r_f = 120$ kpc.}
    \label{fig:eigenfreq}
\end{figure}

We first examine the eigenspectra for each $n, \ell$ combination tested, depicted in figure~\ref{fig:eigenfreq}. We find a densely populated spectrum that points to the shallowness of the gravitational potential. This matches the expected outcomes for galaxies that live in the nonrelativistic limit. Furthermore, we can identify that the lowest energy state is the $n=0, \ell=0$ ground state, as the eigenvalue of this state penetrates the centrifugal barrier and lives within the most gravitationally bound region of the entire galaxy. 

Further qualitative analysis reveals that, over the tested range, a log-linear plot of $1-\omega_{n\ell}$ versus the radial mode $n$ exhibits nonlinear, concave-up behavior for all $\ell$, while a log-log plot of the same parameters appears approximately linear. The nonlinear nature of the log-linear plot indicates that the eigenspectrum we are examining does not follow a geometric trend, whereas the approximate linearity of the log-log plot is consistent with a power law relation. Analysis of this eigenspectrum for greater excited states may be the focus of future work, but the power law trend suggested by the current data is expected by the WKB quantization of bound states in smooth potentials (like the background we introduce in this work).  Finally, a power law relation of the sort $1-\omega_{n\ell} \sim cn^{-k}$ has a beat period of $n^{k+1}$ between neighboring eigenmodes, revealing the rotational timescale of superposed objects like the dark matter halos in this research. 

\begin{figure}[htbp]
    \centering
    \includegraphics[width=1\linewidth]{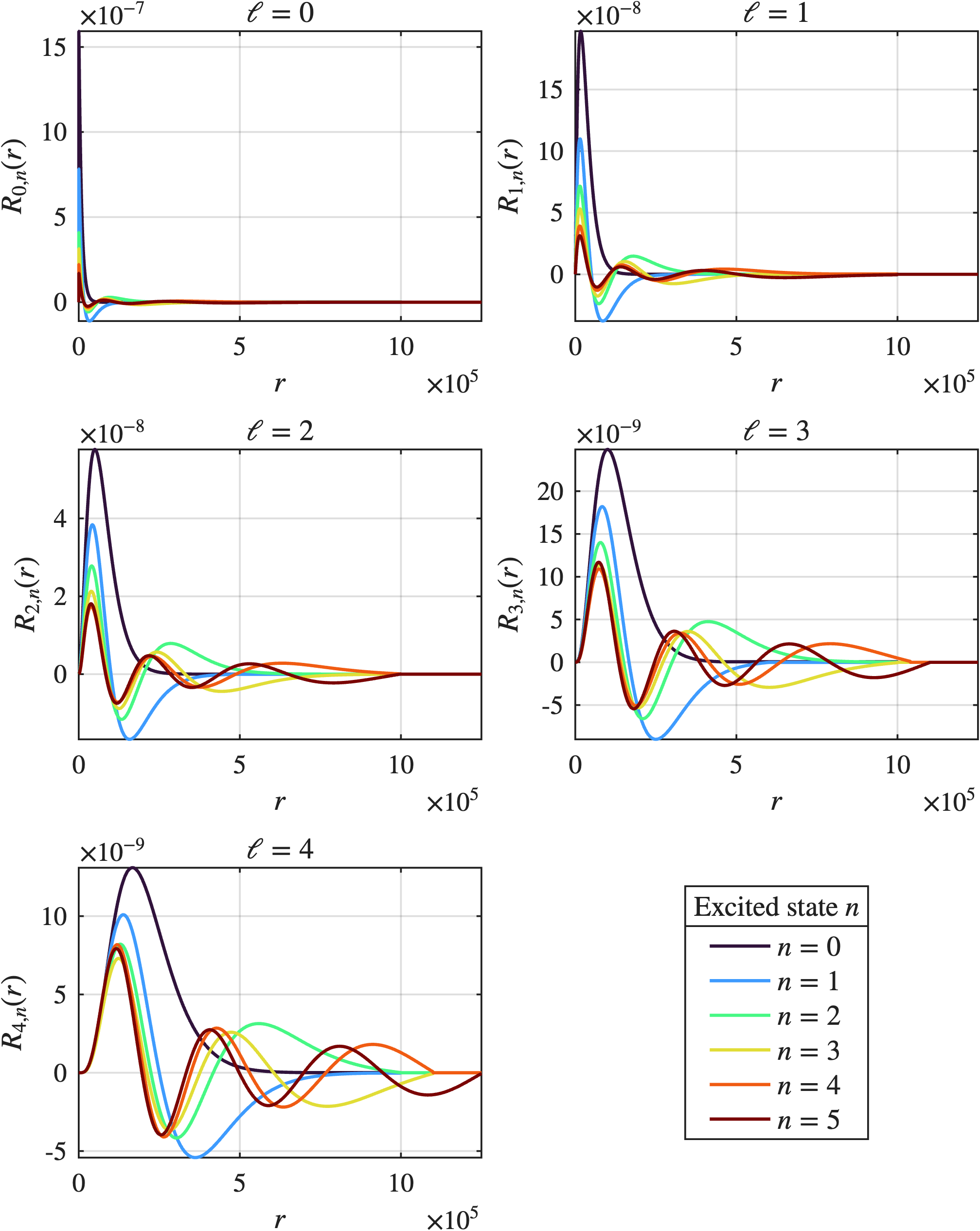}
    \caption{These are the unscaled, natural unit amplitudes and behaviors of the $n = 0,1,2,3,4,5$ excited states for $\ell = 0,1,2,3,4$ as functions of $r$. These radial functions are utilized to create superpositions for the overall halo structure.}
    \label{fig:radial}
\end{figure}

Using the given excited and angular momentum state numbers for a unique eigenfrequency, we can then solve eq. 2.5 as an ordinary differential equation in MATLAB. In doing so, we yield the radial part of the gravitational basis states, and we now have all of the information needed to reconstruct classical ULDM fields. The radial functions for these gravitational eigenmodes are depicted in figure~\ref{fig:radial}. We note that these radial functions and the corresponding interferences that we find from superposing eigenmodes are qualitatively insensitive to the conditions of $r_c$, $r_f$, and $v_0$, provided they match a physical picture of a galaxy.

\subsection{Spiral morphology}
\begin{figure}[htbp]
    \centering
    \includegraphics[width=1\linewidth]{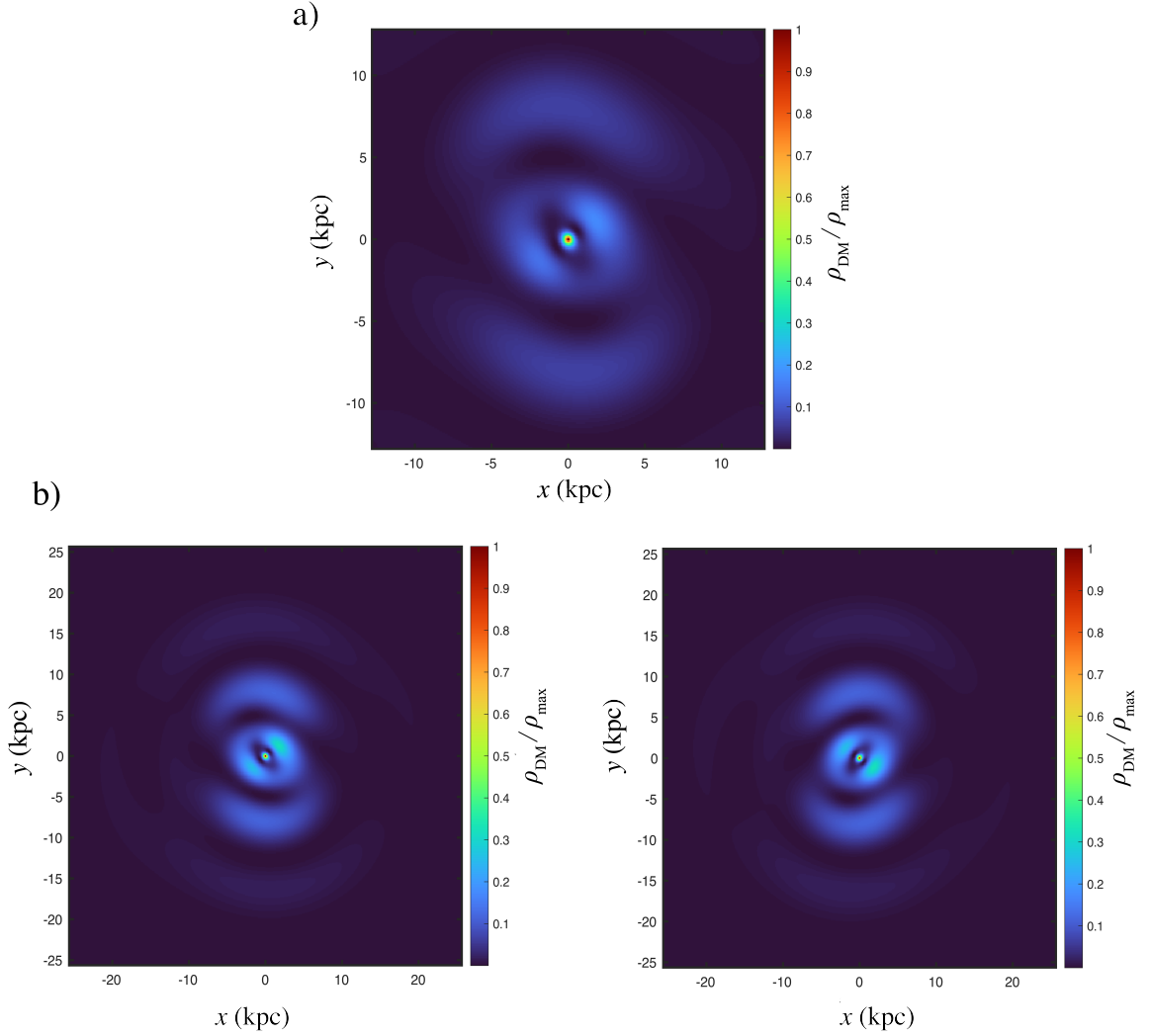}
          \caption{a) We show an exemplary trailing spiral asymmetry in the dark matter halo for a distinct superposition of basis states. This halo is rotating counter-clockwise. b) These two images depict other exemplary spiral forms and natural density waves. We see a trailing and leading spiral within the same period of rotation (left and right respectively) for another superposition of basis states as the halo rotates counter-clockwise. Both superpositions represented are arbitrary constructions which place an emphasis on higher $n$, even $\ell$ states. The exact superpositions can be found in tables~\ref{tab:coeff1} and~\ref{tab:coeff2}.}
    \label{fig:spirals}
\end{figure}

The application of this model to individual dark matter halos has yielded interesting, long-lasting gravitational structures. Through the phase interference of different eigenmodes, we discover dark matter halos which both rotate and break spherical symmetry, along with field solutions which match predicted density wave configurations known to yield spiral patterns in baryonic matter. The structures we find rotate with normal periodicity, and some superpositions yield both trailing and leading arm spirals, as shown in figure~\ref{fig:spirals}. 

\subsection{Physical scaling}
This model suggests that spiral-like patterns in the Universe can arise intrinsically from the nature of dark matter as a wave-like particle. The physical relevance of this model is further supported by scientific data. For the mass of our particle, $m$, we assume the mass of ultralight dark matter (ULDM) which is $\approx10^{-22}$ eV\cite{ferreira2021,hui2017}. The Compton wavelength of such a particle is $\lambda = \frac{\hbar c}{m c^2}$. Using the ULDM mass, we find $\lambda = 0.064$ pc. Scaling the unitless radial parameter in our simulations by this value suggests that the spirals arising through interference of gravitational eigenmodes in the dark matter occur at distances between $10-20$ kpc from the halo center, the same profile of radial size observed in many spiral galaxies. 

Furthermore, the model of dark matter halos we propose rotate at speeds comparable with baryonic matter in galaxies. For most of the gravitational eigenstates superposed, the difference between their energy eigenvalues is on the scale of $\Delta \omega<10^{-8}$. The rotation period of our dark matter halos are governed by $T_{rot} \approx \frac{2\pi}{\Delta \omega}$ in natural units, so to get a physically meaningful quantity we incorporate $T_{\mathrm{rot}} = \frac{2\pi\hbar}{\Delta\omega}$ and $\Delta\omega$ is in terms of the ULDM mass, or $10^{-22}$ eV. We find that $T_{rot} \gtrsim 10^8$ years under such conditions. For comparison, the period of rotation of the Milky Way galaxy is approximately 250 million years\cite{reid2016}. The fact that our dark matter halo model rotates on the same timescale as baryonic matter is a promising outcome which suggests the possibility for baryonic matter to truly take on the shape of these long-lived dark matter spirals. 

Unlike classical physics, the excited states whose interference determines these timescales are long-lived rather than transient. Without known significant electromagnetic or strong couplings, there are extremely restricted thermal de-excitation pathways for dark matter's gravitational eigenmodes to relax through. This suggests that most halos' excited state occupations are determined by primordial initial conditions and significant evolutionary events such that excited states do not decay to a ground state within relevant timescales. In summary, both the size scale and period of rotation of dark matter spirals in this model correspond directly to observable baryonic physics, highlighting the potential for dark matter to give galaxies their long-lived spiral shape, especially when in isolated regions of space.
\subsection{Self-consistency}
\begin{figure} [htbp]
    \centering
    \includegraphics[width=0.75\linewidth]{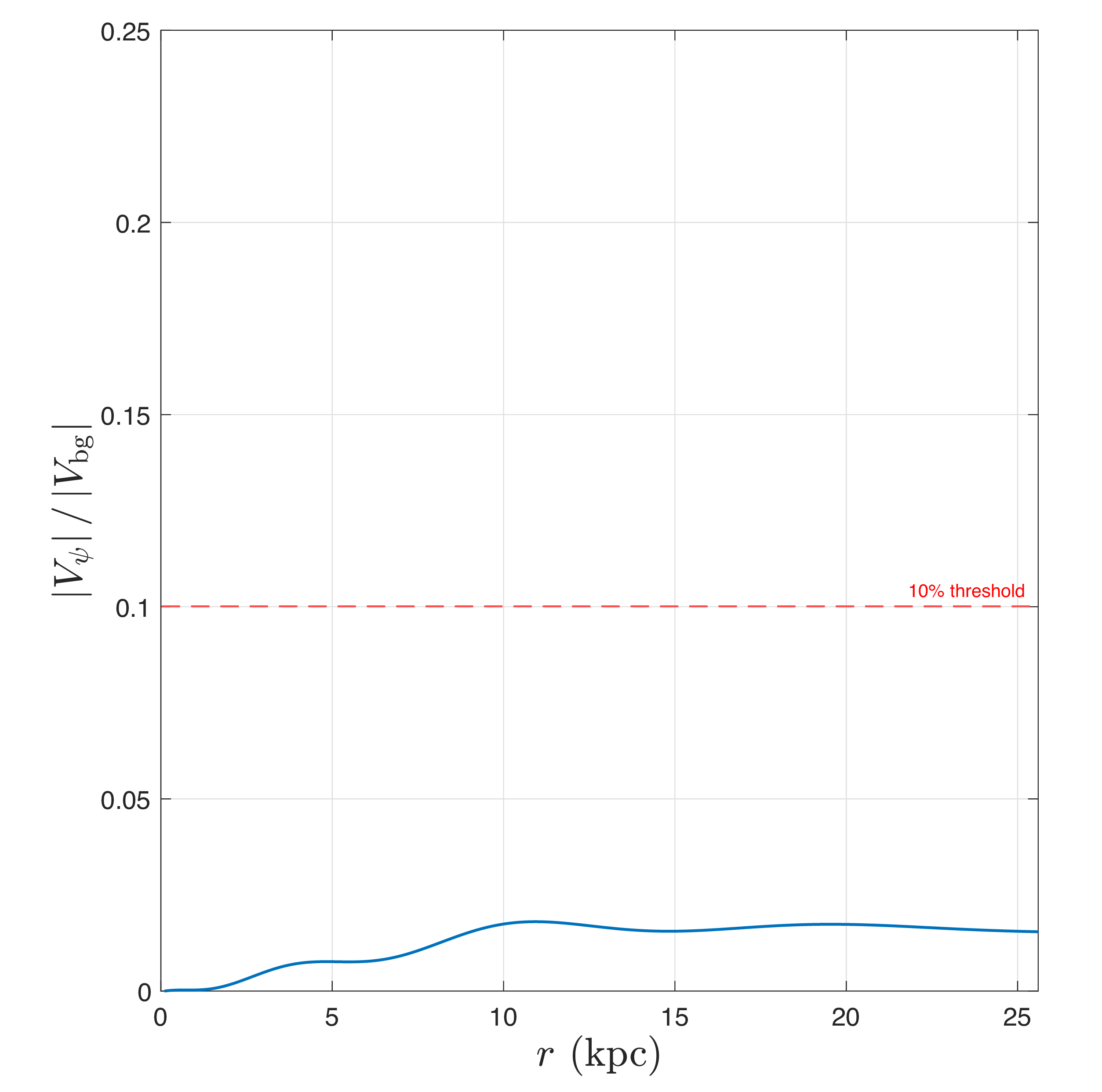}
        \caption{This figure depicts the proportion of azimuthally averaged field potentials attributed to over-densities in dark matter interference compared to the fixed background potential used to calculate individual gravitational eigenmodes. The superposition of modes which generates this plot can be found in table~\ref{tab:coeff3}.}
    \label{fig:backreaction}
\end{figure}

Obtaining the results above necessitates a self-consistency check of our model. In particular, we must verify that the gravitational structures created by superposing eigenmodes solved in a fixed metric do not significantly change the potential in which these gravitational states exist. If the gravitational structures (like the spiral arms) that came out of the fixed potential approximation yielded a potential that was greater than the background, it would be necessary to include backreactions of dark matter over-densities in the dark matter physics itself. In other words, we would need to solve the Schrödinger-Poisson system to incorporate dark matter's self-interactions rather than use our linear approximation based upon a fixed background. While the linearity of the superposed eigenmode fields suggests that any interference patterns may be scalable down to a negligible size using a free parameter, the normalization condition, based upon the overall mass of the dark matter halo, fixes such a parameter instead. To accurately verify the self-consistency of the method, we compute the azimuthally averaged field density profile as a function of radius for different scalar field superpositions. From this, we can calculate the enclosed mass of the local field and compare its gravitational potential to the background potential in which the eigenstates were calculated. The results of this test, plotted in figure~\ref{fig:backreaction}, indicate that local dark matter regions contribute $\sim2\%$ of the total potential, suggesting that our model with a fixed background is sufficient and consistent to approximate the dynamics of these gravitationally bound eigenmode fields throughout both space and time - the backreactions are negligible so long as the field is properly normalized. 

\section{Discussion}

By modeling dark matter halos as large collections of gravitational eigenmodes which satisfy the Klein-Gordon system in a fixed gravitational potential, we are able to show that dark matter may take on spatially asymmetric field configurations. Furthermore, we find solutions which both rotate and form spiral arms in a manner that matches observations of spirals in galaxies. The period of rotation and spatial scale over which interference patterns arise between gravitational eigenmodes suggest the potential for coupling with baryonic matter and forming both trailing and leading spirals in isolated systems. This model also supports previous work which suggests that spirals in galaxies can form naturally from density waves in dark matter halos, an innate feature of this model of dark matter\cite{bray2010}.

Observationally, it is a fact of nature that the majority of spiral galaxies have trailing arms rather than leading arms. Though scientists have found it difficult to quantify the exact number, it is often concluded that $>90 \%$ of observed spiral galaxies have trailing arms\cite{iye2019,pasha1982}. The toy model being proposed is time-symmetric, and therefore does not prefer trailing arms to leading arms in dark matter halos the way that nature appears to prefer trailing arms to leading arms in galactic disks. We suggest that dark matter may be partly responsible for the spiral geometry of galaxies, while baryonic physics - energy dissipation, rotational shearing, and entropy creating processes - are responsible for the preferred trailing orientation of the resulting spiral arms and the time asymmetry seen in galactic physics.

Attempting to incorporate a time asymmetry into the physics of the dark matter itself using classical dissipation, we find that dark matter halos would disappear on non-physical time scales. The introduction of a $+b\partial_t$ to the right side of the Klein-Gordon equation, eq. 2.2, could introduce time asymmetric physics which selects trailing arms over leading ones. However, if we assume that this spiral selection occurs over the course of a single rotational period (such that $b \approx\frac{1}{T_{rot}}$), we must account for the exponential decay of the halo's mass like $M(t) = M(0)\,e^{-bt}$. This leads to a mass half-life of about 0.69 times a single rotational period of the halo, a time scale comparable to the rotational period of a galaxy. This rate of mass loss is not an observationally supported phenomenon, so any future attempt at constructing time-asymmetric dark matter must do so in a new way. This further supports our conclusion that baryonic physics is responsible for the selection of trailing arms over leading arms in the Universe, and it is not the dark matter itself that exhibits this asymmetry.

It is important to emphasize that our model is a simplification of astrophysical systems, and it is currently limited in its ability to predict certain features of spiral galaxies and dark matter halos like pitch angles, arm counts, and features of the large-scale structure of the Universe. Additionally, we note that not all superpositions of eigenmodes form spiral-like patterns - halos which are dominated by low-$\ell$ eigenmodes will reveal a near spherically symmetric dark matter halo, while asymmetries and rotations become more common in halos with a greater population of high-$\ell$ basis states. This suggests that our model is consistent with observations of both ordinary (non-spiral) galaxy models and spiral galaxy features. The proportion of low-$\ell$ to high-$\ell$ states in any superposition is based on the individual history of that halo, accounting for a variety of phenomena that may influence the rotational dynamics of the gravitational eigenmodes like mergers, accretion, and even the Big Bang. As such, this toy model of dark matter halos should be taken as a possible concurrent explanation for spiral patterns in galaxies, particularly isolated spiral galaxies in which tidal events and mergers are not responsible for the development of asymmetric spiral arms in visible matter. This model helps explain a potential mechanism of generic and random spiral development in dark matter halos that may seed, drive, or sustain long-lived spiral patterns in galaxies.  

\section{Conclusion}

We solved the Klein-Gordon equation in a fixed background potential for gravitational eigenmodes when $\ell = 0-4$ and $n = 0-5$. Superposing the resulting functions, we find the possibility to create long-lived spiral-like structures in dark matter halos. These interference dynamics occur on a spatial scale between $10-20$ kpc and have a rotational period of $\gtrsim 10^8$ years, consistent with observed spiral galaxies. The fixed-potential assumption is validated by a backreaction test indicating that density wave perturbations from eigenmode interference in this model account for $\sim 2\%$ of the overall potential. 

Future research should probe the parameter space of our model to quantitatively contrast spiral-like superpositions to non-spirals. Hydrodynamic simulations would be useful to understand how baryonic matter may interact with this model of individual dark matter halos as well. Finally, we suggest the need for observational predictions regarding the possibility of novel pulsar timing array signals and gravitational wave detection\cite{hobbs2010,khmelnitsky2014} based upon the interference of gravitational eigenmodes that this model highlights. 

Ultimately, we show the potential for dark matter halos to take on both trailing and leading arm spiral patterns, and show that individual ULDM halos can naturally provide the long-lived density waves which seed spiral patterns in observable matter.

\acknowledgments

We thank Dr. Arun Kannawadi and Dr. Mark Kruse for their insights into the physics of dark matter and astroparticle physics, as well as Dr. Tom Witelski for his help navigating the numerics for this project. The authors also thank the Department of Mathematics and the Department of Physics at Duke University for their support in conducting this research and the resources which they have provided. 

\paragraph{Code Availability:} The MATLAB R2023 code used to solve the eigenvalue problem and superpose relevant eigenmodes is available from the corresponding author upon reasonable request.

\newpage
\appendix
 
\section{Superposition coefficients}
\label{app:coefficients}
 
These are the exact superpositions which were used to generate the plots in figures~\ref{fig:spirals} and~\ref{fig:backreaction}. Each entry corresponds to a unique coefficient $A_{n\ell}$ for the corresponding eigenmode. The tables show the radial mode index $n$ as columns, and the angular momentum number $\ell$ in the rows. Recall that the field is constructed based upon eq. 2.12.
 
\begin{table}[h]
\centering
\begin{tabular}{|c|c c c c c c|}
\hline
$\ell \,\backslash\, n$ & 0 & 1 & 2 & 3 & 4 & 5 \\
\hline
0 &  2 &  2 &  1 &  1 & 15 & 18 \\
1 &  1 &  1 &  1 &  1 &  0 &  0 \\
2 &  2 & 10 & 10 & 10 & 60 & 30 \\
3 &  1 &  0 &  0 &  0 & 10 & 10 \\
4 & 10 & 15 & 15 & 15 & 40 & 50 \\
\hline
\end{tabular}
\caption{\label{tab:coeff1}
Amplitude coefficients $A_{n\ell}$ for the top panel of figure~\ref{fig:spirals}.}
\end{table}
 
\begin{table}[h]
\centering
\begin{tabular}{|c|c c c c c c|}
\hline
$\ell \,\backslash\, n$ & 0 & 1 & 2 & 3 & 4 & 5 \\
\hline
0 &  0 &  2 &  5 &  5 &  5 & 10 \\
1 &  0 &  2 &  0 &  1 &  2 &  3 \\
2 &  2 & 10 & 20 & 10 & 50 & 50 \\
3 &  0 &  2 &  2 &  4 &  4 & 10 \\
4 &  2 & 20 & 20 & 30 & 40 & 50 \\
\hline
\end{tabular}
\caption{\label{tab:coeff2}
Amplitude coefficients $A_{n\ell}$ for the bottom two panels of
figure~\ref{fig:spirals}.}
\end{table}
 
\begin{table}[h]
\centering
\begin{tabular}{|c|c c c c c c|}
\hline
$\ell \,\backslash\, n$ & 0 & 1 & 2 & 3 & 4 & 5 \\
\hline
0 &  1 &  2 &  1 &  1 & 15 & 18 \\
1 &  1 &  1 &  1 &  1 &  0 &  0 \\
2 &  2 & 10 & 10 & 10 & 60 & 30 \\
3 &  1 &  0 &  0 &  0 & 10 & 10 \\
4 & 10 & 15 & 15 & 15 & 40 & 50 \\
\hline
\end{tabular}
\caption{\label{tab:coeff3}
Amplitude coefficients $A_{n\ell}$ for the superposition shown in
figure~\ref{fig:backreaction}.}
\end{table}

\end{document}